\documentclass[a4paper,fleqn,usenatbib]{mnras}
\usepackage{epsfig}   
\usepackage{amsmath}
\usepackage{mathptmx}
\usepackage{txfonts}
\usepackage[T1]{fontenc}
\usepackage{ae,aecompl}
\usepackage{amssymb}	
\usepackage{scalerel}
\usepackage{longtable}
\usepackage{titlesec}
\usepackage{graphicx,epsfig}
\usepackage{psfrag}

\newcommand{\be}{\begin{equation}}
\newcommand{\e}{\end{equation}}
\newcommand{\bear}{\begin{eqnarray}}
\newcommand{\ear}{\end{eqnarray}}

\def\aj{AJ}

\def\apjs{ApJS}

\def\aap{A\&A}

\def\apjs{ApJS}
\def\apjl{ApJ Letters}

\title[Defining green valley with entropic thresholding] {Tracing the green valley with entropic thresholding}
    
\author[Pandey, B.] {Biswajit Pandey\thanks{E-mail:
    biswap@visva-bharati.ac.in} \\ Department of Physics,
  Visva-Bharati University, Santiniketan, Birbhum, 731235, India\\ }
   
 \date{\today}

 \pubyear{2023}
  
\begin{document}
\label{firstpage}
\pagerange{\pageref{firstpage}--\pageref{lastpage}}      
\maketitle
       
\begin{abstract}
The green valley represents the population of galaxies that are
transitioning from the actively star-forming blue cloud to the
passively evolving red sequence. Studying the properties of the green
valley galaxies is crucial for our understanding of the exact
mechanisms and processes that drive this transition. The green valley
does not have a universally accepted definition. The boundaries of the
green valley are often determined by empirical lines that are
subjective and vary across studies. We present an unambiguous
definition of the green valley in the colour-stellar mass plane using
the entropic thresholding. We first divide the galaxy population into
the blue cloud and the red sequence based on a colour threshold that
minimizes the intra-class variance and maximizes the inter-class
variance. Our method splits the region between the mean colours of the
blue cloud and the red sequence into three parts by maximizing the
total entropy of that region. We repeat our analysis in a number of
independent stellar mass bins to define the boundaries of the green
valley in the colour-mass diagram. Our method provides a robust and
natural definition of the green valley.

\end{abstract}


\titlespacing*{\section}{0pt}{0.85\baselineskip}{0.85\baselineskip}
\section{Introduction}
The luminous component of the matter distribution in the present
universe is primarily represented by the galaxies. The galaxies have a
wide variations in their physical properties. Such broad variations
indicate a range of possible formative and evolutionary
pathways. Classifying the galaxies in the nearby universe based on
their physical properties is a step forward towards understanding
their formation and evolution.

The distribution of several galaxy properties exhibit a distinct
bimodality. The colour is one among them. It is the fundamental
property of a galaxy characterizing its stellar population. The
distribution of galaxy colour is known to be strongly bimodal
\citep{strateva, blanton03, bell1, balogh04, baldry04b}.  The
populations corresponding to the two peaks in the distribution of
galaxy colour are usually referred to as the `blue cloud' (BC) and the
`red sequence' (RS). The two representative populations have
noticeably different physical properties. The differences between the
physical properties of the two populations have been studied in
numerous works \citep{strateva, blanton03, kauffmann03,
  baldry04b}. The galaxies in the blue cloud host younger stellar
populations which are actively star forming. Their morphology is
disk-like and they have a lower stellar mass. Contrarily, the galaxies
in the red sequence are mostly passive and represented by an older
stellar population. They mostly have a bulge-dominated morphology and
higher stellar mass. However, these correlations are not
absolute. Observations show a significant number of ellipticals and
spirals in the blue cloud and the red sequence respectively
\citep{schawinski09, masters10} suggesting a complex evolutionary
history of the galaxies.

The bimodality in the colour distributions of the galaxies is not only
seen in the nearby universe but is also observed for the galaxies at
higher redshifts \citep{bell2, brammer09}. \citet{madau96} shows that
the star formation rate sharply declines after $z=1$ which indicates a
significant evolution in the galaxy properties. The luminosity
function of the galaxies in the red sequence has doubled since $z \sim
1$ \citep{bell2,faber07} indicating an ongoing transition of the
galaxies from the blue cloud to the red sequence. Such transitions may
occur due to a number of different secular and environmental processes
or mechanisms. Some possible physical mechanisms that may be
responsible for such transformations are strangulation \citep{gunn72,
  balogh00}, galaxy harassment \citep{moore96, moore98}, starvation
\citep{larson80, somerville99, kawata08}, ram pressure stripping
\citep{gunn72} and gas expulsion through starburst or AGN
\citep{cox04, murray05, springel05}. The cessation of star formation
may also be driven by a number of other physical processes such as
mass quenching \citep{birnboim03, dekel06}, bar quenching
\citep{masters10} and morphological quenching \citep{martig09}. A
proper understanding of these physical processes and mechanisms and
their roles in producing the observed bimodality is a crucial
requirement for modelling galaxy formation and evolution. The observed
bimodality must be explained by the successful models of galaxy
formation. Various semi-analytic models of galaxy formation have been
used to explain the observed bimodality \citep{menci, driver,
  cattaneo1, cattaneo2, cameron, trayford, nelson, correa19}.

A number of different definitions for the blue cloud and red sequence
have been proposed in the literature. \citet{strateva} propose a
colour cut of $(u-r) = 2.22$ to separate the galaxies in blue cloud
and red sequence. \citet{baldry04a} separate the red and blue galaxies
by fitting a double-Gaussian function to the observed $(u-r)$
colour. The observed colour bimodality is sensitive to the luminosity,
stellar mass and the environment \citep{balogh04, baldry06,
  pandey20}. This advocates the use of other galaxy properties along
with colour to separate the blue cloud and the red sequence. There is
an extensive literature on this topic. Numerous works in the
literature propose to separate the blue cloud and the red sequence in
the colour-magnitude plane \citep{baldry04b, faber07, fritz14},
colour-stellar mass plane \citep{taylor15} or the colour-colour plane
\citep{williams09, arnouts13, fritz14} using different empirical
lines.

The narrow region between the blue cloud and the red sequence is
generally termed as the `green valley'(GV) \citep{wyder07}. The green
valley represents a critical phase in the evolution of galaxies over
cosmic time. The galaxies in the green valley are in a transitional
phase, moving from being actively star-forming to becoming
quiescent. They contain a mix of stellar populations, with both young
and old stars coexisting. They would allow us to understand the
stellar population synthesis, stellar evolution, and the overall
demographics of galaxies. There is no unique evolutionary route for
the galaxies that leads them from the blue cloud to the red sequence
through the green valley. The multiple possible evolutionary pathways
from the blue cloud to the red sequence is a major challenge in
understanding the transitioning galaxies in the green valley.

 Identifying the green valley population is crucial for unraveling the
 complex processes that quench the star formation in galaxies. The
 different existing methods for separating the blue cloud and the red
 sequence are mostly empirical. Galaxies in the green valley are
 intermediate between the blue and the red galaxies and are generally
 regarded as contamination in either sample. It is difficult to define
 the green valley from any of these definitions in a precise
 manner. Determining the exact criteria for classifying galaxies in
 the green valley is quite subjective and vary across different
 studies. \citet{schawinski14} provide a definition of the green
 valley using two empirical lines in the colour-stellar mass
 plane. \citet{bremer18} divide the red, blue and green galaxies using
 three broad colour bins based on the surface density of points in the
 colour-mass plane. \citet{coenda18} define the green valley in the
 (NUV-r) colour-stellar mass diagram using empirical lines and study
 the properties of the transitional galaxies in different
 environments. \citet{eales18} analyze the galaxies selected at
 submillimetre wavelength and argue that the green valley may not
 represent a true third population, but merely a smooth transition
 from the blue cloud towards the red sequence.  \citet{angthopo19}
 propose a definition of the green valley using the $4000\AA$ break
 strength. This definition has been used for a detailed study of the
 stellar populations in green valley galaxies
 \citep{angthopo20}. \citet{pandey} use a fuzzy set theory based
 method to classify the red, blue and green galaxies in the
 SDSS. \citet{das21} and \citet{sarkar22} use this classification to
 study the properties of the green valley galaxies and red spirals in
 different environments. \citet{quilley22} relate the morphology of
 galaxies to their evolution and redefine the green valley using the
 mean colour of Hubble types. \citet{noirot22} use the NUVrK
 colour-colour diagram to identify the blue cloud, green valley and
 the red sequence. More recently, \citet{estrada23} define the green
 valley using the shape of the log(sSFR) distribution and study the
 morphological evolution of the transitional galaxies in the CLEAR
 survey. \citet{brambila23} use empirical lines to define the green
 valley in the SFR-stellar mass plane and explore the roles of
 different environments in the quenching of transitional galaxies.

 Most of the definitions of the green valley are either empirical or
 based on certain user defined parameters, which lack solid
 mathematical justifications. Recently, \citet{pandey23} propose a
 parameter free method to separate the galaxies in the blue cloud and
 the red sequence using Otsu's method for image segmentation. However,
 there are no provision for identifying the green valley in their
 method.

The goal of this work is to present a method for identifying the green
valley based on the entropic thresholding \citep{pun81, kapur85}. The
method would provide an unambiguous definition of the green valley
based on the distribution of galaxies in the colour-stellar mass
plane. The advantage of this method is that the green valley can be
identified solely based on the distribution alone without relying to
any empirical lines or user defined parameters.

The plan of the paper is as follows. We describe the data in the
Section 2, explain the method in Section 3, discuss our results in
Section 4 and present our conclusions in Section 5.

\section{SDSS data}
We use the data from the Sloan Digital Sky Survey (SDSS) \citep{york}.
The SDSS is the largest redshift survey of the nearby Universe. It has
gathered the images and spectra of millions of galaxies in the
universe with unprecedented accuracy. The availability of a large
number of galaxies in the SDSS makes it ideally suited for a
statistical analysis of the galaxy bimodality. We obtain the data from
the SDSS DR16 \citep{ahumada} by using a SQL in the SDSS
SkyServer \footnote{https://skyserver.sdss.org/casjobs/}. We select a
contiguous region of the sky that spans $135^{\circ} \leq \alpha \leq
225^{\circ}$ and $0^{\circ} \leq \delta \leq 60^{\circ}$ in the
equatorial co-ordinates. We extract the information of all the
galaxies in this region that lie within redshift $z < 0.3$ and have
r-band Petrosian magnitude within $ 13.5 \le r_{p} < 17.77$. We
construct a volume limited sample of galaxies from this dataset by
applying a cut $-23 \le M_r \le -21$ to the K-corrected and extinction
corrected $r$-band absolute magnitude. The resulting volume limited
sample lies between the redshift limit $0.041 \le z \le 0.120$ and
contains a total $103984$ galaxies. The stellar masses and the
specific star formation rates (sSFR) of the galaxies in our volume
limited sample are obtained from a catalogue based on the Flexible
Stellar Population Synthesis model
\citep{conroy09}. We utilize the concentration index
  $\frac{r_{90}}{r_{50}}$ \citep{shimasaku01} to assess the
  morphology of galaxies. Here, $r_{90}$ and $r_{50}$ represent the
  radii encompassing $90\%$ and $50\%$ of the Petrosian flux,
  respectively. The values for $r_{90}$ and $r_{50}$ for each galaxy
  are extracted from the photoObj table. Additionally,
  \citet{brinchmann04} provide an emission line classification of
  galaxies using the BPT diagram developed by \citet{baldwin81}. This
  classification is recorded in the `bptclass' variable within the
  galSpecExtra table. We retrieve this classification data for all
  galaxies and identify AGNs based on their `bptclass'. We use a
$\Lambda$CDM cosmological model with $\Omega_{m0}=0.315$,
$\Omega_{\Lambda0}=0.685$ and $h=0.674$ \citep{planck20} for our
analysis.

\section{Method of Analysis}

\begin{figure*}
\resizebox{12.0cm}{!}{\rotatebox{0}{\includegraphics{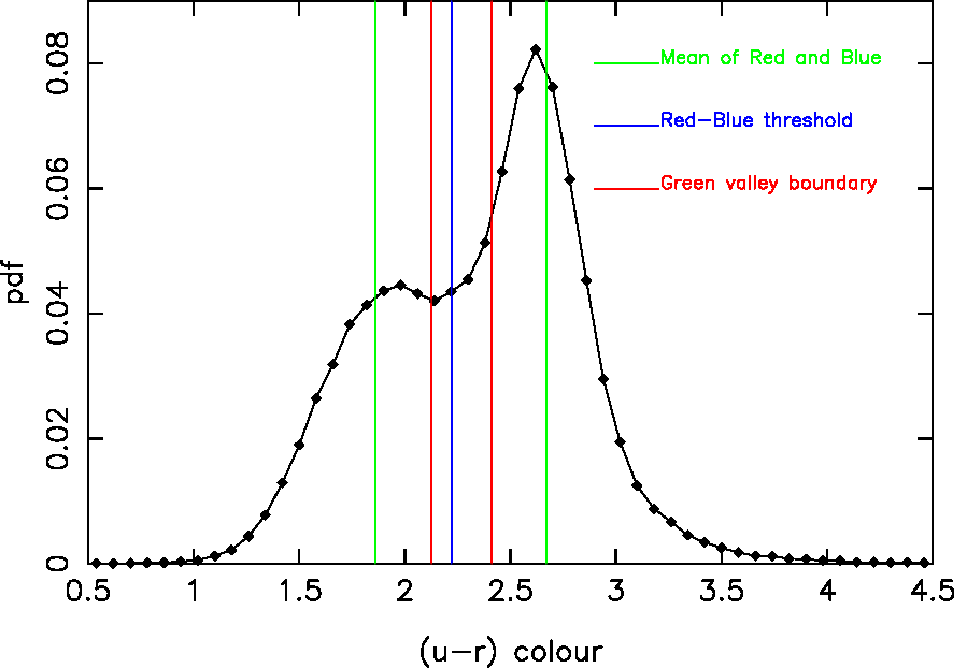}}}
\caption{This figure shows the green valley in the $(u-r)$ colour
  distribution of the entire volume limited sample. We first apply
  Otsu's method to determine an optimal threshold that divides the
  galaxies into two population by minimizing the intra-class variance
  and maximizing the inter-class variance. The intervening region
  between the mean colours of the two populations is then split into
  three parts using the entropic thresholding. The green-valley is
  represented by the region of the PDF bounded by the two red lines in
  the middle. We apply this technique to a number of independent
  stellar mass bins to determine the boundary of the green-valley in
  the colour-stellar mass plane and show it in
  \autoref{fig:entropic}.}
\label{fig:pdf}
\end{figure*}

\begin{figure*}
\resizebox{12.0cm}{!}{\rotatebox{0}{\includegraphics{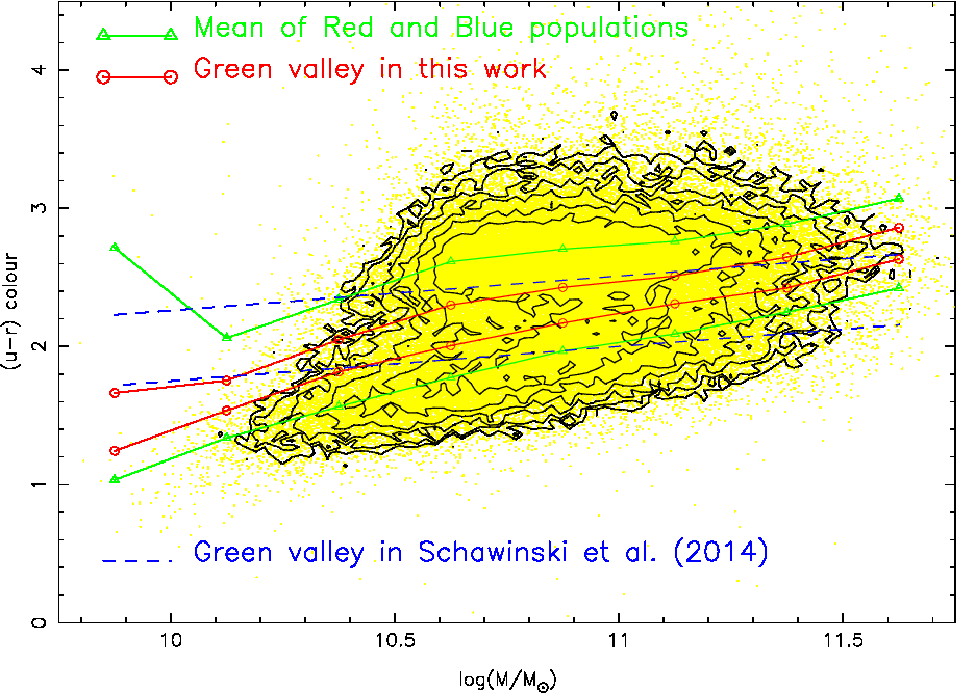}}}
\caption{This shows the distribution of the galaxies
    in the colour-stellar mass plane where each yellow dot represents
    a galaxy. The contours represent regions with different density of
    points. The highest density region is bounded by the innermost
    contour. The boundary of the green valley in this work is shown
    using two solid red lines. We obtain these lines by applying the
    entropic thresholding to a number of independent stellar mass
    bins. The entropic thresholding is applied in the region between
    the mean colours of the blue cloud and the red sequence (the
    region between the two solid green lines) associated with each
    mass bin. The green valley between the two empirical lines
    (blue-dashed lines) defined in \citet{schawinski14} is shown
    together for a comparison.}
\label{fig:entropic}
\end{figure*}


\subsection{Entropic thresholding}
\label{sec:eth}

The information entropy \citep{shannon48} quantifies the uncertainty
in the measurement of a random variable. In other words, it is a
measure of the amount of information necessary to describe a random
variable.  One can define the information entropy $H(X)$ associated
with a discrete random variable $X$ as,

\begin{equation}
H(X) = - \sum^{n}_{i=1} \, p(x_{i}) \, \log \, p(x_{i})
\label{eq:shannon1}
\end{equation}
where $X$ has a total $n$ possible outcomes and $p(x_{i})$ is the
probability of the $i^{th}$ outcome.

In the context of image processing, entropy is a measure of the amount
of information or randomness in an image. It is often used to quantify
the level of uncertainty or disorder in the pixel intensity values
within the image. The high entropy images are more complex with a wide
range of pixel values, whereas the low entropy images are more uniform
and simple. Image segmentation is the process of partitioning an image
into specific regions and extract objects or features of interest. The
image segmentation techniques assume that the object and the
background in the image have different gray-level distributions. One
can divide the pixels into foreground and background by applying
different intensity thresholds and then calculate the sum of their
entropies. The image is optimally thresholded when the sum of the two
class entropies reaches its maximum. Any incorrect separation of the
foreground and background pixels would reduce the total entropy of the
image from its maximum value.

We first describe the idea of entropic thresholding in the context of
image segmentation. \citet{kapur85} propose an algorithm for
distinguishing objects from the background in a gray-level image. In
any thresholding technique, the pixels having intensities greater than
a threshold are identified as a part of the object whereas the
remaining pixels are labelled as the background. One can choose the
threshold intensity based on certain mathematical definitions. The
method proposed by \citet{kapur85} is based on the idea of information
entropy. If N is the total number of pixels in the image and $f_{1},
f_{2},...., f_{n}$ are the gray-level frequencies in $n$ different
bins then the probability of the $i^{th}$ gray-level is given by
$p_{i}=\frac{f_{i}}{N}$. The sum of all these probabilities
$\sum_{i=1}^{n}\,p_{i}=1$ by definition. If the threshold intensity
corresponds to the $s^{th}$ bin then one can define the probability of
the object as,\\
\begin{eqnarray}
  P_{{\scaleto{S}{3.5pt}}}=\sum_{i=1}^{s} p_{i}
\label{eq:Ps}
\end{eqnarray}
The bins $[1,....,s]$ belongs to the object($O$) and the bins
$[s+1,....,n]$ constitute the background($B$). One can derive two
probability distributions corresponding to the two components of the
image as,\\

\begin{eqnarray}
 O: \frac{p_{{\scaleto{1}{3.5pt}}}}{P_{{\scaleto{S}{3.5pt}}}}, \frac{p_{{\scaleto{2}{3.5pt}}}}{P_{{\scaleto{S}{3.5pt}}}},.....\frac{p_{{\scaleto{s}{3.5pt}}}}{P_{{\scaleto{S}{3.5pt}}}}
\label{eq:obj}
\end{eqnarray}

\begin{eqnarray}
 B: \frac{p_{{\scaleto{S+1}{3.5pt}}}}{1- P_{{\scaleto{S}{3.5pt}}}}, \frac{p_{{\scaleto{S+2}{3.5pt}}}}{1- P_{{\scaleto{S}{3.5pt}}}},.....\frac{p_{{\scaleto{n}{3.5pt}}}}{1- P_{{\scaleto{S}{3.5pt}}}}
\label{eq:bkg}
\end{eqnarray}

The information entropy associated with the intensity distribution
of the object is given by,\\
\begin{eqnarray}
  H_{{\scaleto{O}{3.5pt}}}= - \sum_{i=1}^{s} \frac{p_{i}}{P_{{\scaleto{S}{3.5pt}}}} \log \frac{p_{i}}{ P_{{\scaleto{S}{3.5pt}}}} = \log P_{s} + \frac{H_{s}}{P_{{\scaleto{S}{3.5pt}}}}
\label{eq:Ho}
\end{eqnarray}
where
\begin{eqnarray}
  H_{{\scaleto{S}{3.5pt}}} = - \sum_{i=1}^{s} p_{i} \log p_{i}
\label{eq:Hs}
\end{eqnarray}

Similarly, the entropy associated with the intensity distribution of
the background can be expressed as,\\
\begin{eqnarray}
  H_{{\scaleto{B}{3.5pt}}} = - \sum_{i=s+1}^{n} \frac{p_{i}}{1-P_{{\scaleto{S}{3.5pt}}}} \log \frac{p_{i}}{1-P_{{\scaleto{S}{3.5pt}}}} = \log (1-P_{s}) + \frac{H_{n}-H_{s}}{1-P_{s}}
\label{eq:Ho}
\end{eqnarray}
where
\begin{eqnarray}
  H_{n} = - \sum_{i=1}^{n} p_{i} \log p_{i}
\label{eq:Hs}
\end{eqnarray}

The idea is to choose a threshold that maximizes the total entropy
$H_{{\scaleto{O}{3.5pt}}}+H_{{\scaleto{B}{3.5pt}}}$. The same idea can
be extended to any number of objects superimposed on the same
background. This thresholding technique is based on the entropy
maximization principle and is a natural choice in many situations. We
use this thresholding technique for the identification of the green
valley that lies between the blue cloud and the red sequence.

\subsection{Defining the blue cloud and the red sequence using the Otsu's method}

\citet{otsu79} propose a thersholding technique for separating the
foreground and background pixels in a gray-level image, which is
ideally suited for a bimodal distribution of pixel intensities. The
method provides an optimal threshold that minimizes the `intra-class
variance' and maximizes the `inter-class variance'. This method has
been recently used by \citet{pandey23} to separate the galaxies in the
blue cloud and the red sequence. The Otsu's method is a natural choice
for separating the blue cloud and the red sequence from the bimodal
$(u-r)$ colour distribution. We briefly outline the method proposed in
\citet{pandey23}.

We first calculate the probability distribution of $(u-r)$ colour of
the SDSS galaxies in our volume limited sample using $n$ bins. The
probability associated with the $i^{th}$ colour bin is
$p_{i}=\frac{f_{i}}{N}$ where N is the total galaxies in our sample
and $f_{i}$ is the number of galaxies in the $i^{th}$ colour bin.  If
the $(u-r)$ colour threshold corresponds to the $k^{th}$ bin then all
the galaxies in the bins $[1,....,k]$ would belong to the blue cloud
whereas the galaxies in the remaining bins $[k+1,....,n]$ would
represent the red sequence. The probabilities of the class occurrences
for the two populations can be simply written as,\\

\begin{eqnarray}
  P_{{\scaleto{BC}{3.5pt}}}=\sum_{i=1}^{k} p_{i}
\label{eq:w1}
\end{eqnarray}
and
\begin{eqnarray}
  P_{{\scaleto{RS}{3.5pt}}}=\sum_{i=k+1}^{n} p_{i}
\label{eq:w2}
\end{eqnarray}

We iterate through all the possible $(u-r)$ colour thresholds and
estimate the class means for the blue cloud and the red sequence for
each threshold. These are given by,\\

\begin{eqnarray}
  \mu_{{\scaleto{BC}{3.5pt}}}=\frac{\sum_{i=1}^{k} (u-r)_{i} p_{i}}{P_{{\scaleto{BC}{3.5pt}}}}
\label{eq:mu1}
\end{eqnarray}
and
\begin{eqnarray}
  \mu_{{\scaleto{RS}{3.5pt}}}=\frac{\sum_{i=k+1}^{n} (u-r)_{i} p_{i}}{P_{{\scaleto{RS}{3.5pt}}}}
\label{eq:mu2}
\end{eqnarray}
where, $(u-r)_{i}$ is the $(u-r)$ colour corresponding to the $i^{th}$
bin. Clearly, we have
$P_{{\scaleto{BC}{3.5pt}}}+P_{{\scaleto{RS}{3.5pt}}}=1$ for each and
every threshold.

One can also determine the variances in the $(u-r)$ colour of the blue
cloud and the red sequence defined at each threshold as,

\begin{eqnarray}
  \sigma_{{\scaleto{BC}{3.5pt}}}^{2}=\frac{\sum_{i=1}^{k} ((u-r)_{i}-\mu_{{\scaleto{BC}{3.5pt}}})^2 p_{i}}{P_{{\scaleto{BC}{3.5pt}}}}
\label{eq:sigma1}
\end{eqnarray}
and
\begin{eqnarray}
  \sigma_{{\scaleto{RS}{3.5pt}}}^{2}=\frac{\sum_{i=k+1}^{n} ((u-r)_{i}-\mu_{{\scaleto{RS}{3.5pt}}})^2 p_{i}}{P_{{\scaleto{RS}{3.5pt}}}}
\label{eq:sigma2}
\end{eqnarray}

The threshold for the desired separation can be obtained by minimizing
the intra-class variance $\sigma_{intra}^2$ and maximizing the
inter-class variance $\sigma_{inter}^2$.

These can be expressed as,
\begin{eqnarray}
   \sigma_{intra}^2=P_{{\scaleto{BC}{3.5pt}}}\, \sigma_{{\scaleto{BC}{3.5pt}}}^{2}+P_{{\scaleto{RS}{3.5pt}}}\, \sigma_{{\scaleto{RS}{3.5pt}}}^{2}
\label{eq:intra}
\end{eqnarray}
and
\begin{eqnarray}
   \sigma_{inter}^2=P_{{\scaleto{BC}{3.5pt}}}\,P_{{\scaleto{RS}{3.5pt}}}\,(\mu_{{\scaleto{BC}{3.5pt}}}-\mu_{{\scaleto{RS}{3.5pt}}})^2
\label{eq:inter}
\end{eqnarray}

The intra-class and inter-class variances depend on the chosen
threshold. However, their sum
$\sigma_{total}^2=\sigma_{intra}^2+\sigma_{inter}^2$ is independent of
the threshold.

The optimal threshold for the separation of the blue cloud and the red
sequence is the one which simultaneously minimizes the intra-class
variance and maximizes the inter-class variance. \citet{pandey23} show
that this optimal threshold is insensitive to the choice of the number
of bins.

\subsection{Defining the green valley between the blue cloud and the red sequence with entropic thresholding}
The optimal threshold from the Otsu's method divides the galaxies into
the blue cloud and the red sequence but does not help to define the
green valley. The galaxies that are transitioning from the blue cloud
to the red sequence reside near the boundary, and are generally
treated as contamination in either sample. Our primary interest is the
identification of the transitional green valley that must lie
somewhere in the intervening region between the blue cloud and the red
sequence. Consequently, we require to focus only on the intervening
region between the mean colours of the two populations. The class
means $\mu_{{\scaleto{BC}{3.5pt}}}$ and $\mu_{{\scaleto{RS}{3.5pt}}}$
corresponding to the optimal threshold from the Otsu's method
determine the target region for the entropic thresholding. The
galaxies with $(u-r)<\mu_{{\scaleto{BC}{3.5pt}}}$ are undoubtedly a
part of the blue cloud and those with
$(u-r)>\mu_{{\scaleto{RS}{3.5pt}}}$ are definitely a part of the red
sequence.

The traditional entropic thresholding determines a single threshold to
segment an image into two regions (typically foreground and
background). The multilevel entropic thresholding extends this
approach to partition the image into multiple regions based on
different threshold values. This can be particularly useful for
segmenting images with more than two distinct regions. The process of
multilevel entropic thresholding is similar to single-level entropic
thresholding but involves iteratively determining multiple threshold
values. Thus the entropic thresholding also allows one to separate
multiple objects superimposed on a background (\autoref{sec:eth}). The
primary advantage of this method is that it can be applied to
situations where the distributions are not bimodal or multimodal. This
makes it suitable for the identification of the green valley in the
present context. The green valley is in transitional phase and is
intermediate between the blue cloud and the red sequence. In this
work, we are interested to use this technique for tracing the green
valley embedded between the blue cloud and the red sequence. Three
different galaxy populations exist between
$(u-r)=\mu_{{\scaleto{BC}{3.5pt}}}$ and
$(u-r)=\mu_{{\scaleto{RS}{3.5pt}}}$. Our goal is to optimally separate
the three populations in this region. We employ the entropic
thresholding described in \autoref{sec:eth} to split the intervening
region into three parts and distinguish the green valley from the blue
cloud and the red sequence.

The entropy corresponding to the colour distributions of the blue
cloud, the green valley and the red sequence can be respectively
written as,\\

\begin{eqnarray}
  H_{{\scaleto{BC}{3.5pt}}} = \log\Big(\sum_{i=1}^{s_{1}} p_{i}\Big) - \frac{\sum_{i=1}^{s_{1}} p_{i} \log p_{i}}{\sum_{i=1}^{s_{1}} p_{i}}
\label{eq:Hbc}
\end{eqnarray}

\begin{eqnarray}
  H_{{\scaleto{GV}{3.5pt}}} = \log\Big(\sum_{i=s_{1}+1}^{s_{2}} p_{i}\Big) - \frac{\sum_{i=s_{1}+1}^{s_{2}} p_{i} \log p_{i}}{\sum_{i=s_{1}+1}^{s_{2}}  p_{i}}
\label{eq:Hgv}
\end{eqnarray}
and 
\begin{eqnarray}
  H_{{\scaleto{RS}{3.5pt}}} = \log\Big(\sum_{i=s_{2}+1}^{n} p_{i}\Big) - \frac{\sum_{i=s_{2}+1}^{n} p_{i} \log p_{i}}{\sum_{i=s_{2}+1}^{n}  p_{i}}
\label{eq:Hrs}
\end{eqnarray}

We iterate through all the possible thresholds between
$(u-r)=\mu_{{\scaleto{BC}{3.5pt}}}$ and
$(u-r)=\mu_{{\scaleto{RS}{3.5pt}}}$ and choose two thresholds $s_{1}$
and $s_{2}$ (where $s_{1}<s_{2}$) in the interval $[0,n]$ that
maximizes the total entropy $H_{{\scaleto{total}{3.5pt}}} =
H_{{\scaleto{BC}{3.5pt}}}+H_{{\scaleto{GV}{3.5pt}}}+H_{{\scaleto{RS}{3.5pt}}}$.
The two thresholds $s_{1}$ and $s_{2}$ optimally separate the blue
cloud, the green valley and the red sequence. Maximizing the total
entropy ensures that the different classes are optimally
separated. The primary advantage of this method is that it is solely
based on the entropy maximization principle and does not rely on any
user defined relations or parameters.

It may be noted that there are clear relations between the galaxy
colour and the stellar mass or luminosity. So an application of the
method to obtain $s_{1}$ and ${s_{2}}$ from the entire dataset can not
define the green valley in an effective manner.

We apply our method to a number of independent stellar mass bins. We
first calculate $\mu_{{\scaleto{BC}{3.5pt}}}$ and
$\mu_{{\scaleto{RS}{3.5pt}}}$ corresponding to each stellar mass bin
and then apply the entropic thresholding between these $(u-r)$ colours
to obtain the values of $s_{1}$ and ${s_{2}}$ associated with that
bin. This provides us two lines separating the green valley from the
rest of the galaxies in the colour-stellar mass plane. The same method
can be also adopted to define the green valley in the colour-magnitude
plane or stellar mass-SFR plane.

\begin{figure*}
\resizebox{12.0cm}{!}{\rotatebox{-90}{\includegraphics{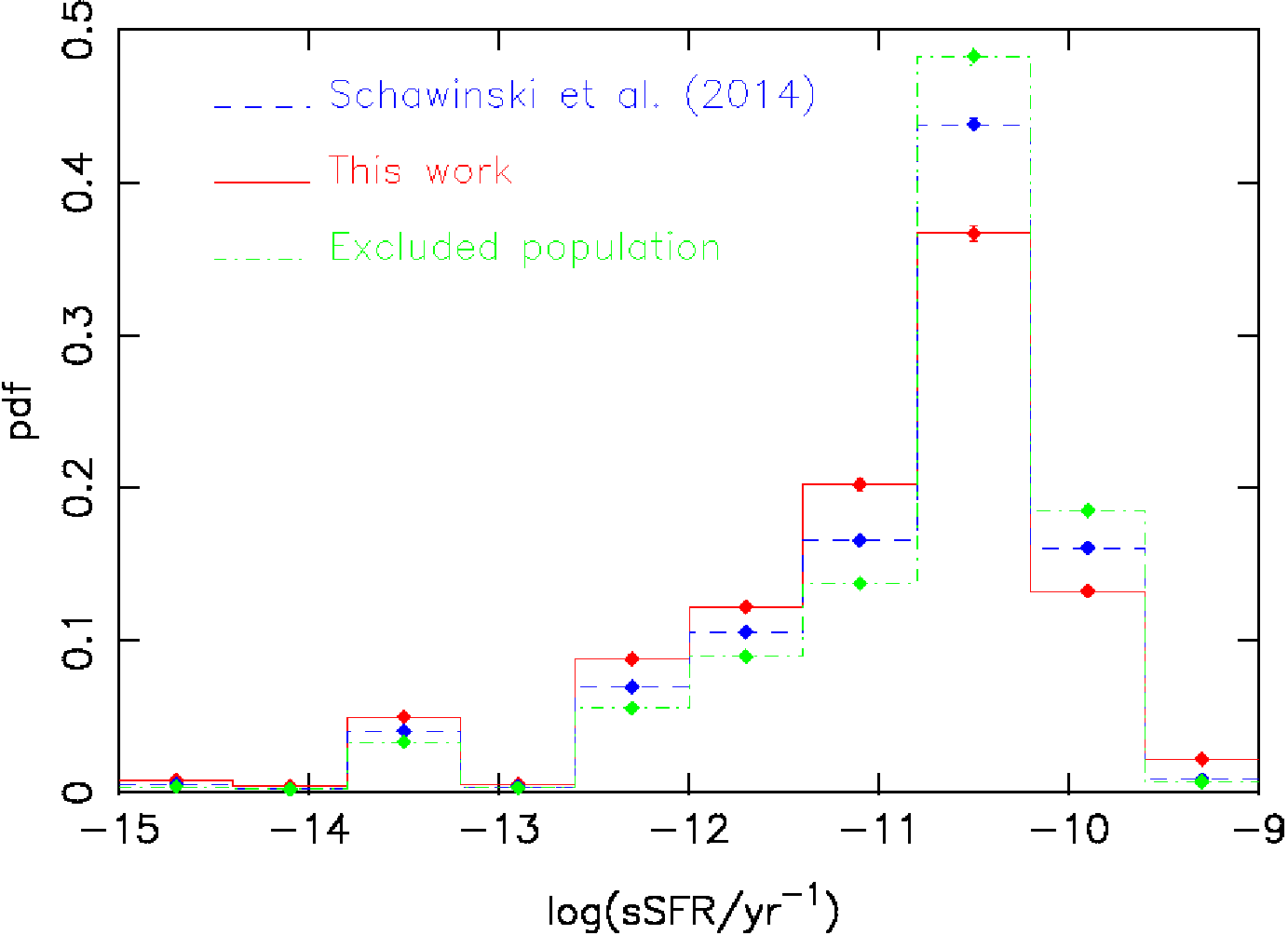}}}
\caption{This shows the distribution of the specific
    star formation rates (sSFR) among the green valley galaxies
    identified through our methodology. We present the sSFR
    distribution for the green valley galaxies that fall between the
    two empirical lines delineated in \citet{schawinski14}, alongside
    the distribution for galaxies excluded by our method, for the
    purpose of comparison. The 1$\sigma$ Poisson errorbars are shown at
    each data point.}
\label{fig:ssfr}
\end{figure*}

\begin{figure*}
\resizebox{12.0cm}{!}{\rotatebox{-90}{\includegraphics{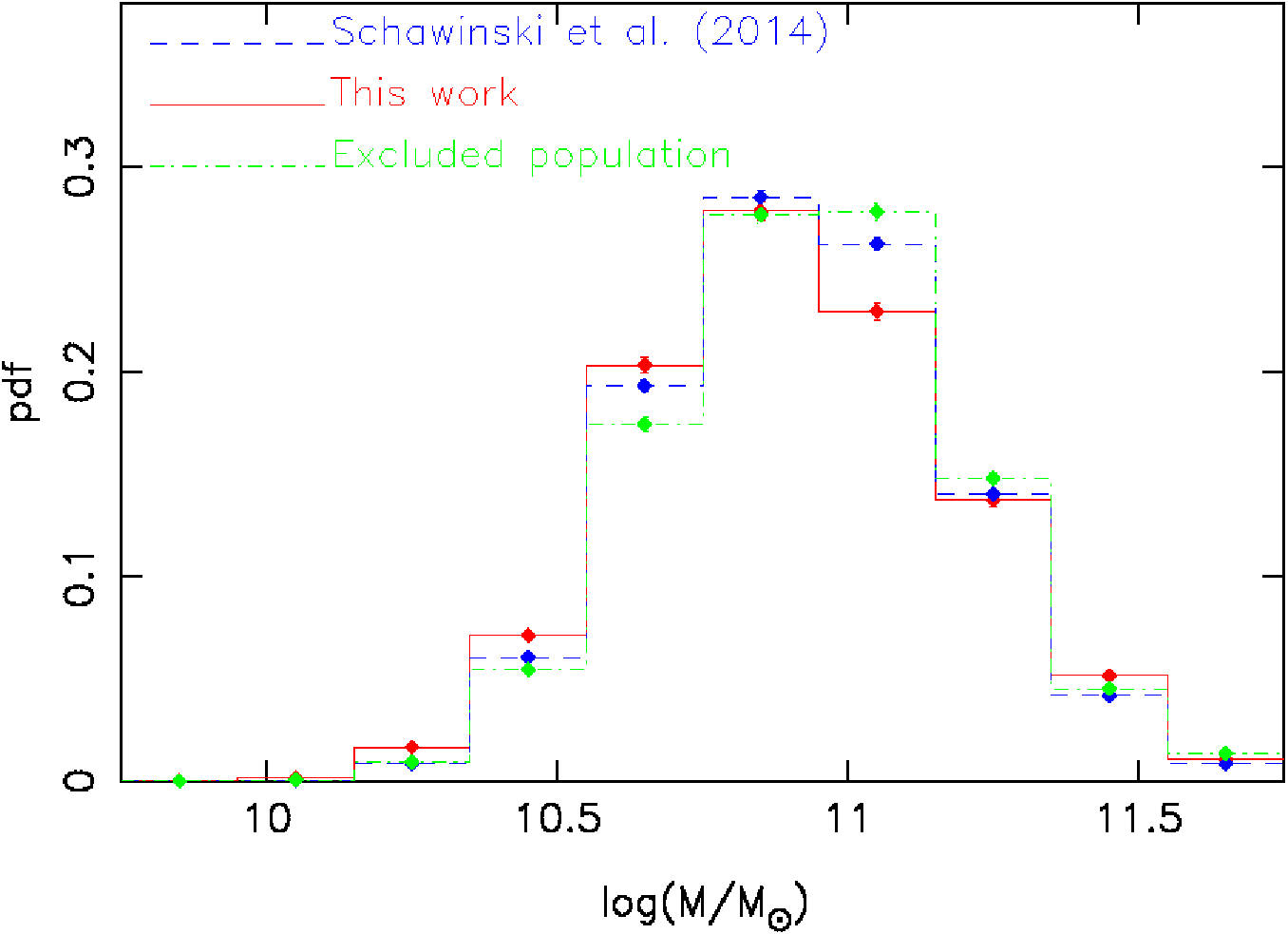}}}
\caption{Same as \autoref{fig:ssfr} but for the stellar mass.}
\label{fig:sm}
\end{figure*}

\begin{figure*}
\resizebox{12.0cm}{!}{\rotatebox{-90}{\includegraphics{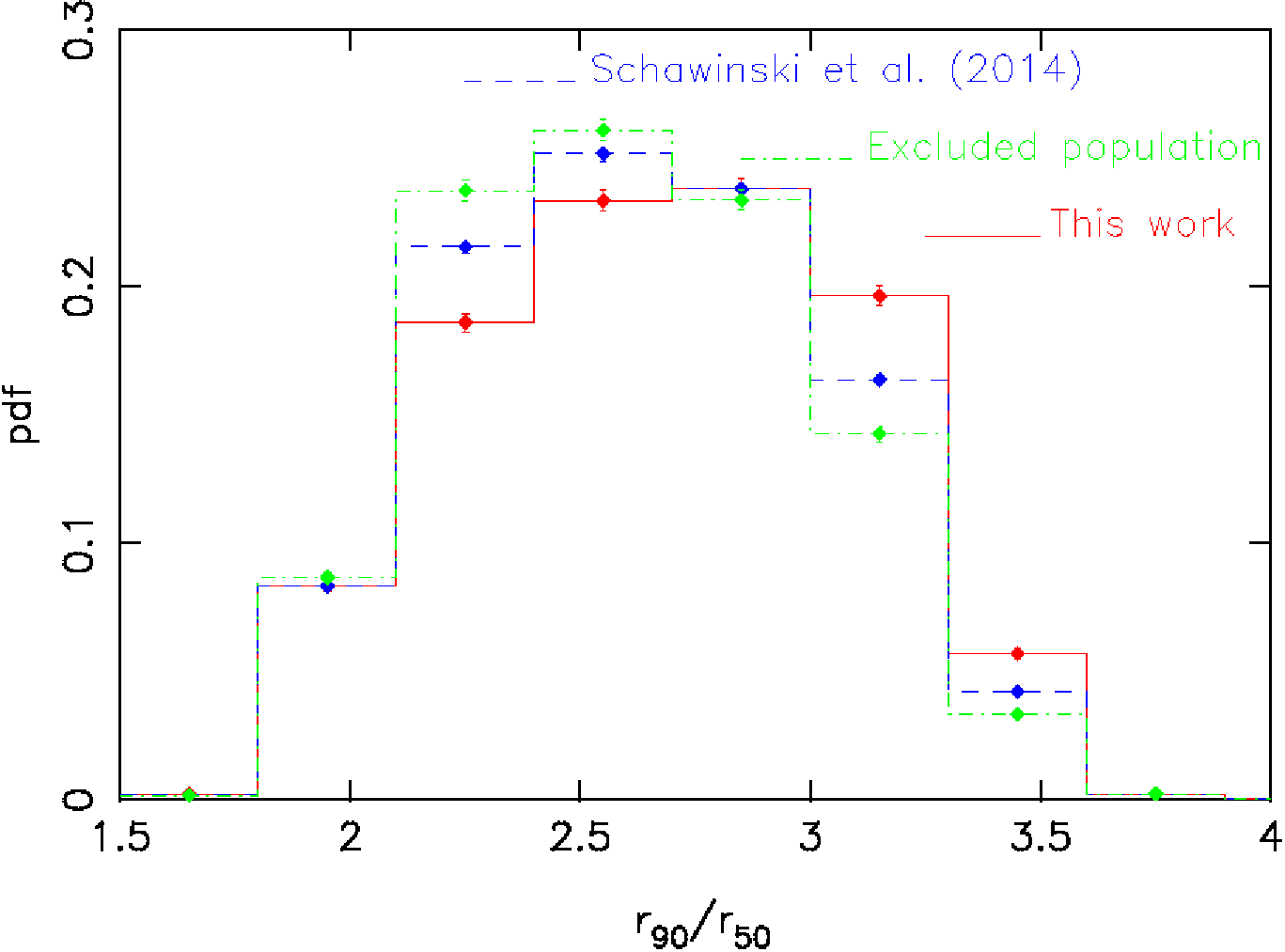}}}
\caption{Same as \autoref{fig:ssfr} but for the concentration index.}
\label{fig:morph}
\end{figure*}

\begin{figure*}
\resizebox{12.0cm}{!}{\rotatebox{-90}{\includegraphics{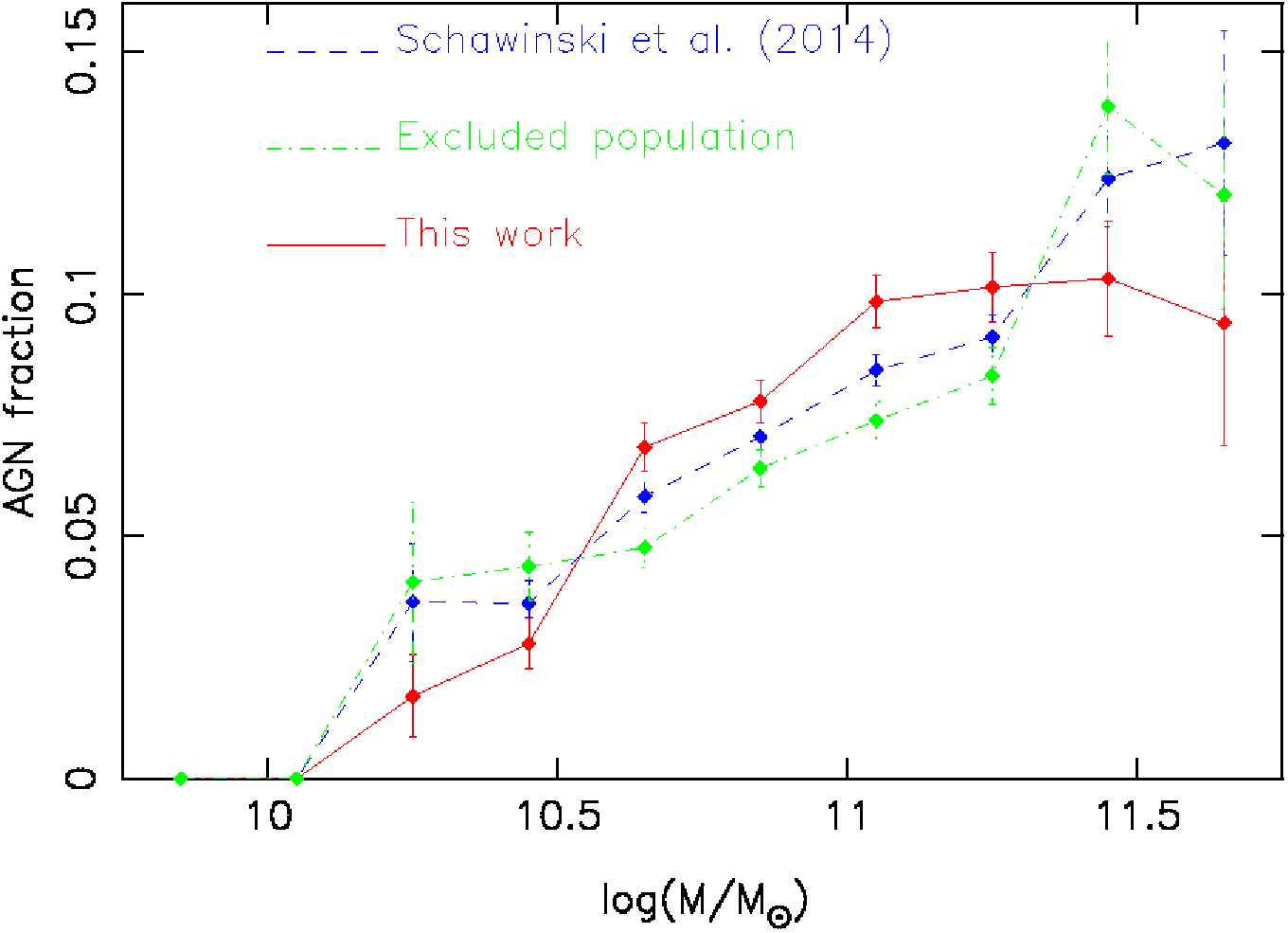}}}
\caption{This figure compares the AGN fraction as a function of
  stellar mass among the green galaxies identified by our method,
  those identified by the empirical method \citep{schawinski14}, and
  those excluded by our method. The 1$\sigma$ Binomial errorbars are
  shown at each data point.}
\label{fig:agn}
\end{figure*}

\section{Results and discussions}

We first demonstrate our method by applying it to the entire dataset
and find out the two colour thresholds that define the green valley in
the $(u-r)$ colour distribution of the galaxies. We calculate the PDF
of the $(u-r)$ colour using $50$ bins and then apply the Otsu's method
to divide the entire distribution into two populations (blue and
red). The Otsu's method provides an optimal colour threshold that
minimizes the intra-class variance and maximizes the inter-class
variance of the two populations. This threshold is insensitive to the
choice of the number of bins \citep{pandey23}. The green valley must
lie in the region between the mean colours of the two populations. The
galaxies with $(u-r)$ colour smaller than the mean colour of the blue
cloud are part of the blue cloud itself. Similarly, the galaxies with
$(u-r)$ colour greater than the mean colour of the red sequence can
only belong to the red sequence. The class uncertainty is only a
characteristic of the intervening region between the mean colours of
the two population. We split this region into three parts using the
entropic thresholding. The three consecutive parts belong to the blue
cloud, the green valley and the red sequence. The green valley is
sandwiched between the blue cloud and the red sequence, which is shown
with two vertical red lines in \autoref{fig:pdf}. It may be noted that
the mean colours of the two populations in \autoref{fig:pdf} seem to
be different from what one would expect by fitting a double-Gaussian
to the colour distribution.  These mean colours are obtained from the
Otsu's method that separates the two populations by minimizing the
within-class variance and maximizing the between-class variance.

 Observations suggest that the colour is strongly correlated with the
 stellar mass. Consequently, two $(u-r)$ colour thresholds are
 incapable to describe the green valley in the entire dataset. We
 repeat our analysis in a number of independent stellar mass bins and
 obtain a pair of colour thresholds corresponding to each mass bin.
 This provides us two boundary lines describing the green valley in
 the colour-stellar mass plane. The green valley is shown with two
 solid red lines in \autoref{fig:entropic}. The two solid green lines
 in this figure represent the variations of the mean $(u-r)$ colours
 of the blue cloud and the red sequence with the stellar mass. The
 number of bins $n$ used for entropic thresholding is a free parameter
 in our method. We repeat our analysis for three different choices of
 $n$ ($n=10$, $n=20$ and $n=30$) and find that the boundary of the
 green valley does not depend on the choice of $n$. This implies that
 our method provides a robust definition of the green valley.
 
 \citet{schawinski14} use two empirical lines to define the green
 valley population in the colour-mass diagram. We show these empirical
 lines (blue dashed lines) in the colour-mass diagram of our data in
 \autoref{fig:entropic} for a comparison. Clearly, the green valley
 defined by the empirical lines of \citet{schawinski14} is much
 broader than the green valley identified by our method. It may also
 be noted that the slopes of the dividing lines in our method change
 with the stellar mass whereas they remain fixed for the empirical
 lines defined in \citet{schawinski14}. These differences may have
 important consequences for any analysis with the green valley
 galaxies. It would be intriguing to compare the star formation rate,
 stellar mass, morphology, and AGN activity of the quiescent galaxies
 identified by our method with those identified by the empirical
 method \citep{schawinski14}.

The application of our method and the empirical method to the same
SDSS dataset yield 14110 and 29184 green valley galaxies,
respectively. We first compare the specific star formation rates
(sSFR) of the green valley galaxies that are identified by the two
methods. The results are shown in \autoref{fig:ssfr}. We find that the
sSSFR distributions of the green valley galaxies identified by both
the methods peak between $10^{-10}-10^{-11}/$yr. Several studies at
low redshift show that the sSFR distribution of the actively star
forming galaxies is skew-lognormal with a peak $\sim 10^{-10}/$yr and
a tail extending towards lower sSFR \citep{wetzel12, eales18}. The low
sSFR tail primarily represents the quiescent galaxies. The actively
star forming galaxies can be separated from the quiescent galaxies by
applying a fixed sSFR boundary at $\log(sSFR/yr^{-1})=-10.5$
\citep{leja22, black22}. This cut shows a mild stellar mass dependence
that indicates that a lower sSFR cut is necessary to separate the
quiescent galaxies from the star-forming galaxies at higher stellar
masses \citep{choi14}. Nonetheless, the galaxies with
$\log(sSFR/yr^{-1})<-10.5$ are mostly quiescent. \autoref{fig:ssfr}
shows that the primary difference between the two green valley
populations identified by our method and \citet{schawinski14} lies in
the abundance of the green valley galaxies near the peak and the tail
of the sSFR distribution. We observe that the empirical method shows a
higher abundance of the green valley galaxies compared to our method
at $\log(sSFR/$yr$^{-1})>-10.5$. On the other hand, there are more
green valley galaxies at $\log(sSFR/$yr$^{-1})<-10.5$ in our method
compared to the empirical method. A higher abundance of the green
valley galaxies at $\log(sSFR/$yr$^{-1})<-10.5$ indicates that our
method is somewhat better at identifying the quiescent population.  A
greater abundance of the green valley galaxies at
$\log(sSFR/$yr$^{-1})>-10.5$ in the empirical method perhaps shows
more interloping of the green valley from the actively star-forming
galaxies in the blue cloud. Our method excludes a significant number
of galaxies from the green valley as identified by the empirical
method. Additionally, we analyze the specific star formation rate
(sSFR) distribution of the excluded population, as illustrated in
\autoref{fig:ssfr}. It is evident that the excluded population
contains a higher proportion of actively star-forming galaxies and a
lower proportion of quiescent galaxies compared to the green valley
population identified by our method. The disparities between the
empirical method and our approach are primarily driven by the excluded
population. The 1$\sigma$ Poisson error bars displayed at each data
point emphasize the statistical significance in the sSFR differences
of the green valley populations identified by the two methods.

The green valley defined in our method and the empirical method have
little to no overlap at lower ($\log \frac{M}{M_{\odot}}<10.2$) and
higher masses ($\log \frac{M}{M_{\odot}}>11.4$)
(\autoref{fig:entropic}).  It may introduce some differences in the
stellar mass distribution of the green valley populations in the two
methods. The stellar mass distribution of the green valley galaxies in
the two methods are compared in \autoref{fig:sm}. It shows that the
stellar mass distribution of the green valley galaxies peaks at
$\log\frac{M}{M_{\odot}}\sim 10.85$ for both the methods. However, the
empirical method exhibits a broader peak compared to our method. The
tails of the stellar mass distribution extend to similar masses in
both methods. However, compared to our method, the empirical method
yields a greater fraction of green galaxies near the peak and a lower
fraction of green galaxies near the tails of the stellar mass
distribution. It suggests that the empirical definition is less
sensitive in identifying the quiescent galaxies at lower and higher
masses. We also examine the stellar mass distribution of the green
valley population that is excluded by our method
(\autoref{fig:sm}). The differences between the stellar mass
distributions of the excluded population and the population identified
by our method are somewhat more pronounced, although they exhibit
similar trends. The 1$\sigma$ Poisson error bars depicted at each data
point highlight statistically significant differences in the stellar
mass distributions of these populations within the intermediate mass
range.

We compare the morphology of green valley galaxies as identified by
the empirical method, our method, and the population excluded by our
method in \autoref{fig:morph}. We observe that the distributions of
the concentration index $\frac{r_{90}}{r_{50}}$ are similar across all
three methods. However, we note statistically significant differences
in the distribution amplitudes within specific ranges: there is a
notably higher amplitude between $3<\frac{r_{90}}{r_{50}}<3.6$ and a
lower amplitude between $2.2<\frac{r_{90}}{r_{50}}<2.6$ in the green
valley identified by our method compared to the other two populations.
It is established that $\frac{r_{90}}{r_{50}}=2.3$ signifies a pure
exponential profile \citep{strateva}, while
$\frac{r_{90}}{r_{50}}=3.33$ describes a pure de-Vaucouleurs profile
\citep{blanton01}. Therefore, a higher concentration index is
typically associated with ellipticals and bulge-dominated systems,
whereas a lower value ($<2.6$) is associated with disk-dominated
spiral galaxies \citep{strateva}. This suggests that the green valley
population identified by our method consists of a higher proportion of
bulge-dominated systems and a lower proportion of disk-dominated
systems compared to both the empirically defined population and the
excluded population.

Several observational studies suggest that AGNs may play a crucial
role in quenching star formation in green valley galaxies
\citep{nandra07,cimatti13, zhang21}. We compare the AGN fraction as a
function of stellar mass for green valley galaxies identified by our
method and the empirical method in \autoref{fig:agn}. The results for
the excluded population are also presented in the same
figure. Notably, compared to the other two populations, the green
valley identified by our method exhibits a statistically significant
higher fraction of AGNs in the mass range $10.5<\log
\frac{M}{M_{\odot}}<11.3$. However, the AGN fraction is relatively
lower at smaller and higher masses for the green valley galaxies in
our method compared to the other two populations. Nevertheless, the
differences are not statistically significant due to relatively larger
error bars at low and high masses.

These comparisons do not show the superiority of our method in
an absolute sense. Nevertheless, our analysis shows the resulting
differences in the physical properties of the green valley galaxies
arising out of their selection.

\section{Conclusions}

We propose a method for identifying the green valley galaxies and
apply it to the SDSS data. The proposed method is based on two
different methods of image segmentation \citep{kapur85, otsu79}. The
resulting green valley in our method occupies a lesser area in the
stellar mass-colour plane compared to the empirical method. The slope
of the boundary lines of the green valley in our method changes with
the stellar mass, whereas they remain the same at all masses in the
empirical approach. 

The comparison of sSFR distributions between green valley populations
identified by our method and the empirical method reveals notable
differences in the abundance of galaxies near the peak and tail of the
distribution. Our method tends to identify more quiescent galaxies at
lower sSFR values, suggesting its effectiveness in distinguishing
between actively star-forming and quiescent populations. The
differences are even higher for the empirically defined green valley
galaxies that are excluded by our method. The analysis of the stellar
mass distribution further emphasizes differences between green valley
populations identified by different methods. While both methods
exhibit similar peak locations, the empirical method yields a broader
distribution with a higher fraction of green galaxies near the peak
and a lower fraction near the tails compared to our method. The
differences are somewhat larger for the population excluded by our
method. We also observe disparities in the morphology of the green
valley populations between the two methods. Our method suggests a
greater prevalence of bulge-dominated systems and a reduced occurrence
of disk-dominated systems compared to both the empirically defined
population and the excluded population. Our method identifies a higher
fraction of AGNs in the intermediate mass range and a lower AGN
fraction at lower and higher masses. The differences in AGN fractions
between populations are not statistically significant at low and high
masses, suggesting potential complexities in the relationship between
AGN activity and green valley galaxies.

The observed differences have some implications on the evolutionary
history of the quiescent galaxies in the green valley. The galaxies
may quench star formation following different evolutionary routes. The
physical processes/mechanisms responsible for quenching are different
in higher and lower mass galaxies.

The galaxies are gas fed from the cosmic streams \citep{dekel06} and
the circumgalactic medium \citep{maller04}. A fraction of the gas
reservoir is converted into stars over every dynamical time. The star
formation decays with age as the gas reservoir is depleted over
time. Besides this natural ageing, galaxies can reach a quiescent
stage due to the physical processes/mechanisms that either prevent the
accumulation of gas or prevent gas from forming stars or expel gas
from the galaxy. A host of physical processes/mechanisms have been
proposed for quenching in the green valley. However, it is difficult
to correctly assess the relative roles of these processes/mechanisms
in quenching the galaxies. A precise identification of the green
valley can provide valuable insights into several aspects of galaxy
evolution. The study of the green valley galaxies would reveal the
different physical processes and quenching mechanisms that are
responsible for the suppression of star formation in the transitional
galaxies.

The green valley population identified by our method occupies a
narrower region in the colour-stellar mass plane. The extent and size
of the green valley in the stellar mass-colour plane can offer
insights into the quenching timescales. Galaxies spending more time in
the green valley may undergo a gradual quenching process, while those
traversing it swiftly may experience a rapid cessation of star
formation. A smaller area of the green valley may suggest a shorter
quenching duration, reflecting the effectiveness of processes
transitioning galaxies from star-forming to quiescent states. This
implies a faster transition through this phase, possibly indicating
more efficient or intense mechanisms driving the halt of star
formation, such as gas depletion or feedback mechanisms. Galaxies
undergoing strong interactions or mergers with other galaxies might
experience rapid cessation of star formation due to gas influx
triggering intense starbursts followed by swift depletion. Likewise,
galaxies hosting active galactic nuclei (AGN) might encounter feedback
processes that promptly suppress star formation. Several earlier
studies \citep{ferrarese00, haring04, kauffmann09, banerjee23} have
identified the presence of a bulge and gas availability as crucial
requirements for AGN activity. Our analysis reveals that the green
valley population identified by our method tends to host a higher
proportion of bulge-dominated systems and a greater fraction of
AGN. This suggests that both merger events and AGN activity may play
significant roles in suppressing star formation in green valley
galaxies, thereby reducing the quenching timescale.

A caveat in our analysis is that we identify the green valley using
only optical colour. A number of studies suggest that the (NUV-r) or
UV-optical colours are superior for the identification of the green
valley \citep{wyder07, salim14}. The primary advantage of the UV over
optical is that it can detect even a low level of ongoing star
formation. We plan to use different colours to identify the green
valley with our method and carry out an in depth study of the green
valley galaxies in a future work.

Finally, we note that the green valley does not have a universally
accepted definition. Different studies define the green valley using
empirical lines in the colour-mass, colour-magnitude or mass-SFR
plane. The exact criteria for identifying the green valley remains
subjective. Keeping this subjectivity in mind, we propose a new
definition of the green valley using the entropic thresholding. The
entropic thresholding is based on the idea of maximum entropy which is
more general in nature. The boundary of the green valley in our method
is exclusively decided by the data. We conclude that our method
provides a natural and robust definition of the green valley.

\section{ACKNOWLEDGEMENT}
The author thanks an anonymous reviewer for the valuable comments and
suggestions that helped to improve the draft. BP would like to
acknowledge the financial support from the SERB, DST, Government of
India through the project CRG/2019/001110. The author thanks Suman
Sarkar for the help with the SDSS data.

Funding for the SDSS and SDSS-II has been provided by the Alfred
P. Sloan Foundation, the Participating Institutions, the National
Science Foundation, the U.S. Department of Energy, the National
Aeronautics and Space Administration, the Japanese Monbukagakusho, the
Max Planck Society, and the Higher Education Funding Council for
England. The SDSS website is http://www.sdss.org/.

The SDSS is managed by the Astrophysical Research Consortium for the
Participating Institutions. The Participating Institutions are the
American Museum of Natural History, Astrophysical Institute Potsdam,
University of Basel, University of Cambridge, Case Western Reserve
University, University of Chicago, Drexel University, Fermilab, the
Institute for Advanced Study, the Japan Participation Group, Johns
Hopkins University, the Joint Institute for Nuclear Astrophysics, the
Kavli Institute for Particle Astrophysics and Cosmology, the Korean
Scientist Group, the Chinese Academy of Sciences (LAMOST), Los Alamos
National Laboratory, the Max-Planck-Institute for Astronomy (MPIA),
the Max-Planck-Institute for Astrophysics (MPA), New Mexico State
University, Ohio State University, University of Pittsburgh,
University of Portsmouth, Princeton University, the United States
Naval Observatory, and the University of Washington.

\section{Data Availability}
The data underlying this article are available in
https://skyserver.sdss.org/casjobs/. The datasets were derived from
sources in the public domain: https://www.sdss.org/

\bsp	
\label{lastpage}
\end{document}